\documentclass[12pt]{iopart}

\usepackage{graphicx}
\usepackage{bm}
\begin{document}
\title{Elastic theory for the vortex-lattice melting in iron-based
high-T$_c$ superconductors}
\author{Qing-Hu Chen$^{1,2}$, Qing-Miao Nie$^{3}$, Jian-Ping Lv$^{1,2}$, Tin-Cheung  Au Yeung$^{1,4}$}

\address{
$^{1}$ Center for Statistical and Theoretical Condensed Matter
Physics, Zhejiang Normal University, Jinhua 321004, P. R. China \\
$^{2}$ Department of Physics, Zhejiang University, Hangzhou 310027,
P. R. China\\
$^{3}$ Department of Applied Physics, Zhejiang University of
Technology, Hangzhou 310023, China\\
$^{4}$ School of Electric and Electronic Engineering, Nanyang
Technological University, Singapore 639798}

\ead{qhchen@zju.edu.cn}

\begin{abstract}
The vortex-lattice melting transitions  in two typical iron-based
high-Tc superconductor $Ba(Fe_{1-x}Co_{x})_{2}As_{2}$ (122-type)
and$Nd(O_{1-x}F_{x})FeAs$ (1111-type) for  magnetic fields both
parallel and perpendicular to the anisotropy axis are studied within
the elastic theory. Using the parameters from experiments, the
vortex-lattice melting lines in the H-T diagram are located
systematically by various groups of Lindemann numbers. It is
observed that the theoretical results for the vortex melting  on
both $Ba(Fe_{1-x}Co_{x})_{2}As_{2}$ and $Nd(O_{1-x}F_{x})FeAs$ for
parallel fields agree well with the recent experimental data. The
future experimental results for the vortex melting can be compared
with the present theoretical prediction by tuning reasonable
Lindemann numbers.
\end{abstract}


\maketitle

\section{Introduction}

Recently, the newly discovered iron-based superconductors have
attracted considerable scientific interest both experimentally
\cite{Kamihara,wen,chen,ren1,ren3,liu,wang,rotter,Sefat,Li,Yuan,122,1111,Ying,1111b,1111c,Hoffman}
and theoretically \cite{Mazin, Si,Xiang,Fang,dft,Zhang}. First, the
novel superconductivity in 1111 phase $FeAs$ superconductors was
reported experimentally, giving a new path to high temperature
superconductivity. The $LaOFeAs$ under doping with $F^{-}$ irons at
the $O^{2-}$ sites was found to exhibit superconductivity with
T$_{c}=26$K ~\cite{Kamihara}, and the superconductivity was also
observed  with holes doping ~\cite{wen}. Then T$_{c}$ was
surprisingly increased up to above 40K when La in
$LaO_{1-x}F_{x}FeAs$ was substituted by other rare earth
elements~\cite {chen,ren1}. It was quickly observed later that
T$_{c}$ is about 55K in $SmO_{1-x}F_{x}FeAs$ ~\cite{ren3,liu} and
$Gd_{1-x}Th_{x}OFeAs$ ~\cite{wang}.  It is a first non-copper-based
superconductors in which the maximum critical temperature is much
higher than the theoretical value predicted from BCS theory
\cite{mc}. On the other hand, the 122 phase iron-based
superconductor $BaFe_{2}As_{2}$ was discovered more recently
\cite{rotter}, and the superconducting critical temperature  was
found to be as high as $38$K by hole-doping.  It is observed that
the electron doping of which by Co \cite{Sefat} and Ni ~\cite{Li}
also induce superconductivity. Although there are some theoretical
studies on iron-based superconductors, the type-II superconductivity
as well as the mechanism of supeconductivity  are not well
understood to date.

The vortex-lattice solid (glass with random pinning) state with
zero linear resistivity is crucial for the application of
high-$T_{c}$ superconductors, thus the melting of vortex-lattice
in bulk type-II superconductors is of great
significance~\cite{brandt,Houghton,Safar,cubitt,Herbut}. The main
aspect of Lindemann criterion suggests that the lattice melts when
the root mean square thermal displacements of the components of a
lattice reach a certain fraction of the equilibrium lattice
spacing, such a criterion was first adopted to study the
vortex-lattice melting transition in type-II superconductors with
a magnetic field parallel to the anisotropic axis, then this
approach was used to draw the melting lines in the case of
magnetic field perpendicular to the anisotropic
axis~\cite{blatterG,Calson,huchen,nie}. Since the upper critical
field is also high\cite{1111,122} in iron-based high-Tc
superconductors, the thermal fluctuation may drive  the
vortex-lattice to a vortex liquid in a field far below  the upper
critical field \cite{blatterG} through vortex melting.  We will
extend the elastic theory  to study the vortex melting  in these
newly discovered superconductors.

In this paper, using the parameters measured in recent
experiments, we study the vortex-lattice melting transitions for
two typical iron-based layered superconductors
$Ba(Fe_{1-x}Co_{x})_{2}As_{2}$ (122-type) with  low anisotropy
~\cite{Sefat,122} and $Nd(O_{1-x}F_{x})FeAs$ (1111-type) with high
anisotropy \cite{ren1,1111}, in the framework of the elastic
theory. The melting lines for  magnetic fields both parallel and
perpendicular to the anisotropic axis are systematically located
with different groups of Lindemann numbers. A comparison with
current experimental findings is made. The present paper is
organized as follows.  In Section 2, we introduce the theoretical
method used in this work, Section 3 presents the main results,
finally, we give a short summary.

\section{Elastic theory}

We will consider the field $B$ both parallel and perpendicular to
the anisotropy axis (i.e. c-axis in this paper), the elastic
theories in both cases are presented respectively.

\subsection{Thermal fluctuations}

Whether the field is parallel or perpendicular  to c-axis, for
ideal triangular vortex-line lattice,  the free energy in elastic
theory can be expressed in an unified way with quadratic terms of
the deviation vector $\mathbf{u}=(u_x,u_y)$ describing the
fluctuations of vortices from their equilibrium
positions\cite{brandt,Houghton,blatterG,Calson,huchen,nie}
\begin{equation}
F=\frac 12\int\limits_{}\frac{d^3\mathbf{k}}{(2\pi
)^3}\mathbf{u}\cdot \mathbf{C}\cdot \mathbf{u},
\end{equation}
The matrix $\mathbf{C}$ for  fields parallel to the c-axis is
different from that for fields perpendicular to it. We denote
$\mathbf{C^c}$ and $\mathbf{C^{ab}}$ to be  the elastic matrix for
the fields parallel  and  perpendicular to c-axis, respectively,
which are given as follows
\begin{equation}
\mathbf{{C^c}=\left(
\begin{array}{cc}
c_Lk_x^2+c_{66}k_{\perp }^2+c_{44}k_z^2 & c_Lk_xk_y \\
c_Lk_xk_y & c_Lk_y^2+c_{66}k_{\perp }^2+c_{44}k_z^2
\end{array}
\right) }
\end{equation}
and
\begin{equation}  \label{e.2}
\mathbf{{C^{ab}}=\left(
\begin{array}{cc}
c_{11}k_{x}^{2}+c_{66}^{h}k_{y}^{2}+c_{44}^{h}k_{z}^{2} &
c_{11}k_{x}k_{y}
\\
c_{11}k_{x}k_{y} &
c_{66}^{e}k_{x}^{2}+c_{11}k_{y}^{2}+c_{44}^{e}k_{z}^{2}
\end{array}
\right)}
\end{equation}

In matrix  $\mathbf{C^{c}}$, $k_{\perp }^2=k_x^2+k_y^2$,
$c_{66},c_L$, and $c_{44}$ are the wave-vector-dependent shear,
bulk, and tilt elastic moduli\cite{Houghton,brandt1,campell},
respectively, which are determined as follows

\begin{eqnarray}
c_{44}(\mathbf{k}) &=&\frac{B^2}{4\pi }[\frac M{M_z}]
\frac{1-b}{2b\kappa ^2}[\frac 1{k_{\perp
}^2+(M/M_z)(k_z^2+m_\lambda^2)}+1]
\end{eqnarray}
and
\begin{eqnarray}
c_{11}(\mathbf{k}) &=&\frac{B^2}{4\pi }\frac{1-b}{2b\kappa
^2}[\frac{k^2+(M/M_z)m_\lambda^2}{
(k^2+m_\lambda^2)[k_{\perp }^2+(M/M_z)(k_z^2+m_\lambda^2)]}  \nonumber \\
&&-\frac 1{k_{\perp }^2+(M/M_z)k_z^2+m_\xi ^2}]
\end{eqnarray}
Where we have defined $m_\lambda ^2=\frac{1-b}{2b\kappa ^2} $ and
$m_\xi ^2=\frac{1-b}{b}$, $M$ is a quasi-particle effective mass
in $xy$-plane, $M_z$ describes along c-axis, $b=B/B_{c2}$ with
$B_{c2}$ is the upper critical field, and $\kappa =\lambda _{\perp
}/k_{\perp }$. The bulk modulus is $c_L=c_{11}-c_{66}$ with shear
modulus
\begin{equation}
c_{66}=\frac{B_{c2}^2}{4\pi }\frac{b(1-b)^2}{8\kappa ^2}.
\end{equation}

The matrix $\mathbf{C^{ab}}$ also contains some elastic moduli.
Since the anisotropy exists here, tilt and shear modulus are not
isotropic any more. For example, $c_{44}^{h}(\mathbf{k})$ ($
c_{44}^{e}(\mathbf{k})$) is tilt modulus along (perpendicular to)
c-axis.  Similarly, $c _{66}^{h}$ ($c _{66}^{e}$) represents shear
modulus parallel to (perpendicular to) the anisotropy axis $ c$.
For the detailed expressions for elastic moduli, one may refer to
Ref. \cite{nie}

The thermal fluctuations of the vortices are given by inverting the
kernel $  \mathbf{C(k)}$ as follows
\begin{eqnarray}
\langle u_\alpha ^2\rangle  &=&\frac{k_BT}{(2\pi )^3}\int
d\mathbf{k}[\mathbf{C}^{\beta}_{\alpha\alpha}]^{-1}(\mathbf{k}),
\alpha =x,y,  \beta =c,ab
\end{eqnarray}

The integrations in Eq. (7) are over $0<k_z<\infty$ and within the
first Brillouin zone (BZ) for $k_x$ and $k_y$. To be specific, we
consider a lattice structure as the low temperature equilibrium
state as shown in Fig. 1 of Ref. \cite{nie}, where the lattice
spacings are $a_{x} = a$ and $a_{y} = a\sqrt{3/2}$ for fields
parallel to c-axis, $a_{x} = a/\sqrt{\gamma}$ with
$\widehat{x}\parallel \widehat{c}$ and $a_{y} = a\sqrt{3\gamma}/2$
for fields perpendicular to c-axis with $\gamma ^{2}=m_{c}/m_{ab}$
and $a=\sqrt{{2}\Phi_{0}/{\sqrt{3}B}}$.

For parallel fields, we consider the mean-square displacement of a
vortex lattice  from the equilibrium $ d^2(T)=u_x^2+u_y^2$ which can
be written as \vspace{0.0cm}

\begin{eqnarray}
d^2(T)=\frac{k_BT}{(2\pi )^3}\int
d\mathbf{k}[\frac{1}{c_{66}k^2_{\perp}+c_{44}k^2_z}+\frac{1}{c_{11}k^2_{\perp}+c_{44}k^2_z}]
\end{eqnarray}

For convenience, we introduce the dimensionless wave vector ${\bf
q}=(k_{x}/\Lambda _{x},k_{y}/\Lambda _{y},k_{z}/\Lambda )$, where
\(\Lambda_{x,y}\) are the wave numbers at the edges of the first BZ.
Explicitly, they are given as $\Lambda_{x} = {4\pi}/{3a}$ and
$\Lambda_{y} ={2\pi}/{a\sqrt{3}}$ for fields parallel to c-axis and
$\Lambda_{x} = {4\pi\sqrt{\gamma}}/{3a}$ and $\Lambda_{y}
={2\pi}/{a\sqrt{3\gamma}}$ for fields perpendicular to c-axis. The
unit  of wave number in \(k_z\) direction is taken as
$\Lambda=\sqrt{{4\pi B}/{\Phi _{0}}}=\sqrt{{ 8}\pi/{\sqrt{3}a^2}}$.
The thermal fluctuations are then expressed as
\begin{eqnarray}
d^2(T)= \frac{\Lambda }{B^{2}}\frac{k_{B}T}{(2\pi )^{3}} \int d{\bf
q} \widetilde{C}_{\alpha\alpha}^{-1}({\bf q}\equiv
\frac{\Lambda}{B^{2}}\frac{k_{B}T}{(2\pi )^{3}} I_{dd}, {\;\;\;} for
B||c
\end{eqnarray}
and
\begin{eqnarray}
\label{e.4} \langle u_{\alpha }^{2}\rangle & = & \frac{\Lambda
}{B^{2}}\frac{k_{B}T}{(2\pi )^{3}} \int d{\bf q}
\widetilde{C}_{\alpha\alpha}^{-1}({\bf q}) \equiv
\frac{\Lambda}{B^{2}}\frac{k_{B}T}{(2\pi )^{3}} I_{\alpha\alpha},
{\;\;\;} \alpha= x,y, for B \perp c
\end{eqnarray}
with the normalized matrix $\widetilde{{\bf C}}({\bf q})\equiv {\bf
C}({\bf q})/B^{2}\Lambda_x \Lambda_y$.

\subsection{Lindemann criterion }
The Lindemann criterion presumes that the lattice melts, when the
root mean square thermal displacements of the components of a
lattice reach some fraction of the equilibrium lattice spacing.
For two cases, we write the usual isotropic Lindemann criterion
for parallel fields,
\begin{equation}
\label{e.26} \langle d^{2}\rangle =c^{2}a^{2}.
\end{equation}
and the anisotropic one for perpendicular fields,
\begin{equation}
\label{e.26} \langle u_{x}^{2}\rangle =c_{x}^{2}a_x^{2},{\;\;\;}
\langle u_{y}^{2}\rangle =c_{y}^{2}a_y^{2}.
\end{equation}
with $c_x$ and $c_y$ are two Lindemann numbers for two transverse
directions. Combining the Lindemann criterion and the elastic
theory, we can get the melting equations for two cases,
\begin{equation}
\label{e.27} \frac{t}{\sqrt{1-t}}
=c^2\frac{32\pi^4\kappa^2}{\sqrt{3\varepsilon_d}}
\frac{\sqrt{b}}{I_{dd}},{\;\;} for B||c
\end{equation}
and
\begin{equation}
\label{e.27} \frac{t_\alpha}{\sqrt{1-t_{\alpha}}} =
(\frac{a_{\alpha}}{a})^2
c_{\alpha}^2\frac{32\pi^4\kappa^2}{\sqrt{3\varepsilon}}
\frac{\sqrt{b_{\alpha}}}{I_{\alpha\alpha}},{\;\;}\alpha=x,y, for B
\perp c
\end{equation}
with $\varepsilon$ the Ginzburg parameter $\varepsilon_d= 16 \pi^3
\kappa^4(k_BT_c)^2/\Phi_0^3H_{c2}^{c}(0)$ and $\varepsilon= 16 \pi^3
\kappa^4(k_BT_c)^2/\Phi_0^3H_{c2}^{ab}(0)$ .

\subsection{Layer pinning effect }

When fields perpendicular to c-axis, the effect of layer pinning
reduces fluctuations in both directions and induces an additional
momentum-independent term to the elastic matrix in Eq. (3) such that
\begin{equation}
\label{e.25} {\bf C^{ab}_{lp}}=\left(
\begin{array}{cc}
c_{11}k_{x}^{2}+c_{66}^{h}k_{y}^{2}+c_{44}^{h}k_{z}^{2}+\Theta &
c_{11}k_{x}k_{y} \\
c_{11}k_{x}k_{y} &
c_{11}k_{y}^{2}+c_{66}^{e}k_{x}^{2}+c_{44}^{e}k_{z}^{2}
\end{array}
\right),
\end{equation}
where
\begin{equation}
\label{e.23} \frac{\Theta}{\Lambda ^{2}B^{2}}=\frac{4\sqrt{\pi
}}{\beta _{A}\kappa ^{2}\gamma}(\frac{\xi
_{c}^{0}}{s})^{3}\frac{1-b}{b^{2}}\frac{1}{(1-t)^{3/2}}e^{-\frac{8}{1-t}(
\frac{\xi _{c}^{0}}{s})^{2}}
\end{equation}
is proportional to the critical depinning current ~\cite{Ivlev},
with $\xi _{c}^{0}\equiv \xi _{c}(T=0)$, $\beta_A \approx 1.16$ and
$s$  the layer separation.

\section{Results}

\subsection{Vortex melting in $Ba(Fe_{1-x}Co_{x})_{2}As_{2}$.}

We study the vortex melting in  $Ba(Fe_{1-x}Co_{x})_{2}As_{2}$
(x=0.1) as a representative of 122-type iron-based
superconductors\cite{Yuan}. The parameter measured from a most
recent experiment  \cite{122} gives   $T_{c}=22$K,
$\lambda_{\bot}=160nm$, $\xi_{\bot}=2.44nm$,
$H_{c2}^{c}(T=0)=50$T, $H_{c2}^{ab}(T=0)=70$T, and $s=23A$. The
anisotropic parameter $\gamma$ falls from 2.0 to 1.5 with the
decrease of temperature.  In our calculation, we observe that the
melting lines determined in the framework of elastic theory only
change  slightly with the variation of $\gamma$ (in the range of
1.5 $\sim$ 2.0). So we set $\gamma=2$ in our calculations.

\begin{figure}
\includegraphics[width=0.8 \textwidth,height=10.cm]{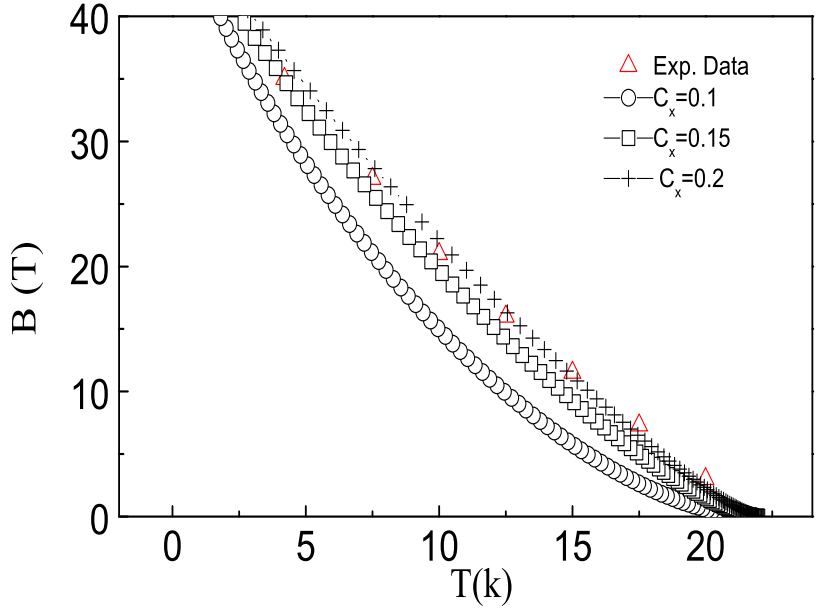}
\caption{\label{Figure Graph1}  Comparison of vortex-lattice melting
lines for  $Ba(Fe_{1-x}Co_{x})_{2}As_{2}$  with experimental results
in Ref. ~\cite{122} for magnetic fields parallel to c-axis. }
\end{figure}

We first calculate the melting line for the  parallel fields  with
various Lindemann numbers. It is interesting  to note in  Fig. 1
that the vortex melting line for $Ba(Fe_{1-x}Co_{x})_{2}As_{2}$
obtained in this paper is well consistent with  the irreversible
line measured experimentally \cite{122} with Lindemann number
$c=0.2$. The irreversibility line in superconductor is usually
regarded as the melting line. \cite{blatterG}

\begin{figure}
\includegraphics[width=1.0 \textwidth,height=8.cm]{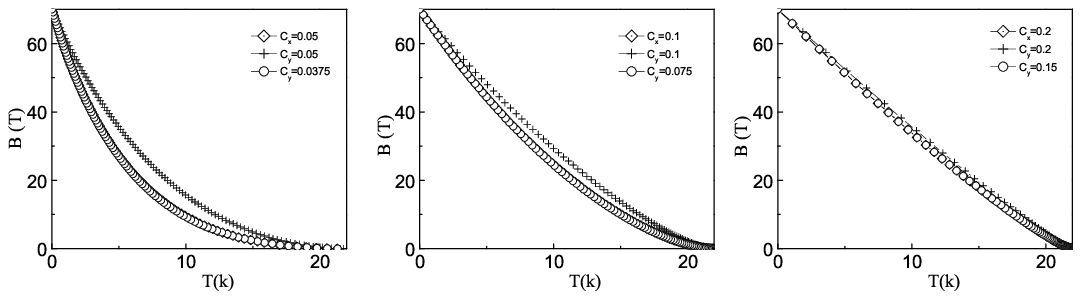}
\caption{\label{Figure Graph1}  Vortex-lattice melting lines for
 $Ba(Fe_{1-x}Co_{x})_{2}As_{2}$ with   magnetic field
perpendicular to c-axis (without intrinsic layer pinning). From
left to right, $C_{x}=0.05,C_{y}=0.05,C_{y}=0.0375$;
$C_{x}=0.1,C_{y}=0.1,C_{y}=0.075$;
$C_{x}=0.2,C_{y}=0.2,C_{y}=0.15$. }
\end{figure}

Then we study  the vortex melting for perpendicular fields. Due to
the laking of the experimental data until now, three groups of
Lindemann numbers are selected to calculate the melting lines
($C_{x}=0.05,C_{y}=0.05,C_{y}=0.0375$;
$C_{x}=0.1,C_{y}=0.1,C_{y}=0.075$;
$C_{x}=0.2,C_{y}=0.2,C_{y}=0.15$).  The obtained melting curves
are shown in Fig. 2. The future experimental results can be
inserted into this figure, and be compared with  interpolation
values obtained here.

Setting $C_{x}=C_{y}$, we obtain two curves as shown in Figs.
2(a-c), as in Ref. \cite{Calson}.  As pointed out in
Ref.~\cite{huchen}, to interpret these two curves as two " melting
lines " and thus reach to a conclusion of an intermediate phase is
unphysical, since the elastic theory can give at best one melting
line.  To impose the same Lindemann number along the two directions
is of no physical basis. In order to achieve a single melting line,
one can tune the Lindemann numbers in the two directions. A good
collapse of the melting lines in two directions can be achieved if
setting the ratio $ c_x/c_y \approx 1.33$. It is interesting to note
that this ratio is very close to $1.37$ observed in Ref. \cite{nie}
using parameters of cuprate superconductors.

The vortex melting for fields perpendicular to the c-axis is also
influenced by the layer pinning. In order to study this intrinsic
layer pinning  effect,  the matrix (15) is used to calculate the
thermal fluctuations along two transverse directions. The melting
lines for two groups of Lindemann number
($C_{x}=0.1,C_{y}=0.1,C_{y}=0.075$;
$C_{x}=0.2,C_{y}=0.2,C_{y}=0.15$) are collected in  Fig. 3. A good
collapse can also be obtained by setting the same ration $ c_x/c_y
\approx 1.33$.  The future experimental data can be also compared
with these values.

\begin{figure}
\includegraphics[width=0.8 \textwidth,height=10.cm]{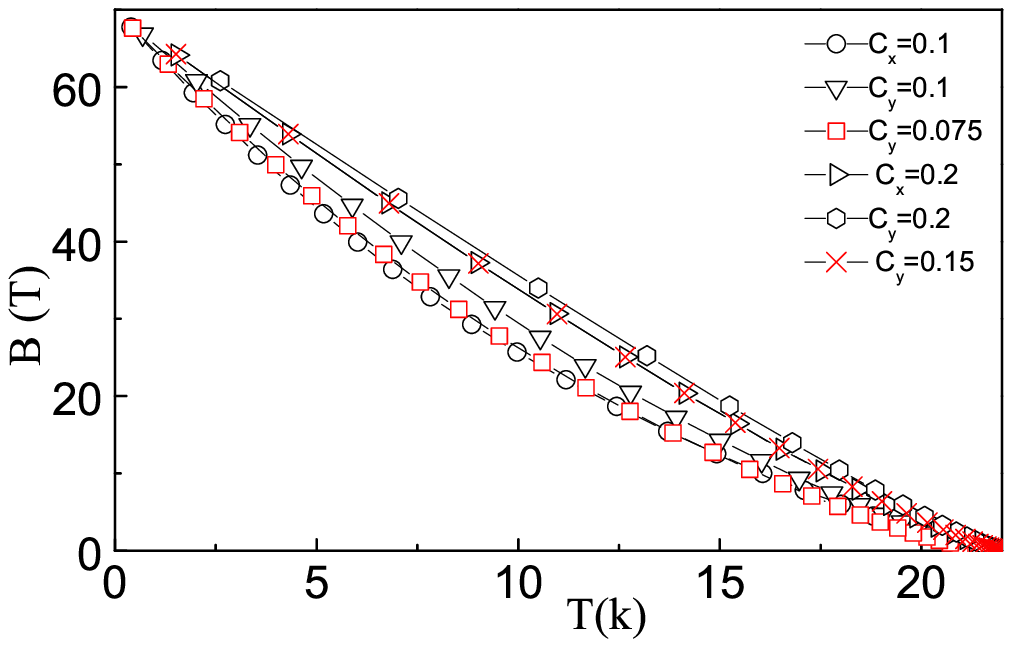}
\caption{\label{Figure Graph1}  Vortex-lattice melting lines for
 $Ba(Fe_{1-x}Co_{x})_{2}As_{2}$ with   magnetic field
perpendicular to c-axis (with  intrinsic layer pinning).}
\end{figure}

\subsection{Vortex melting in $Nd(O_{1-x}F_{x})FeAs$.}

\begin{figure}
\includegraphics[width=0.8 \textwidth,height=10.cm]{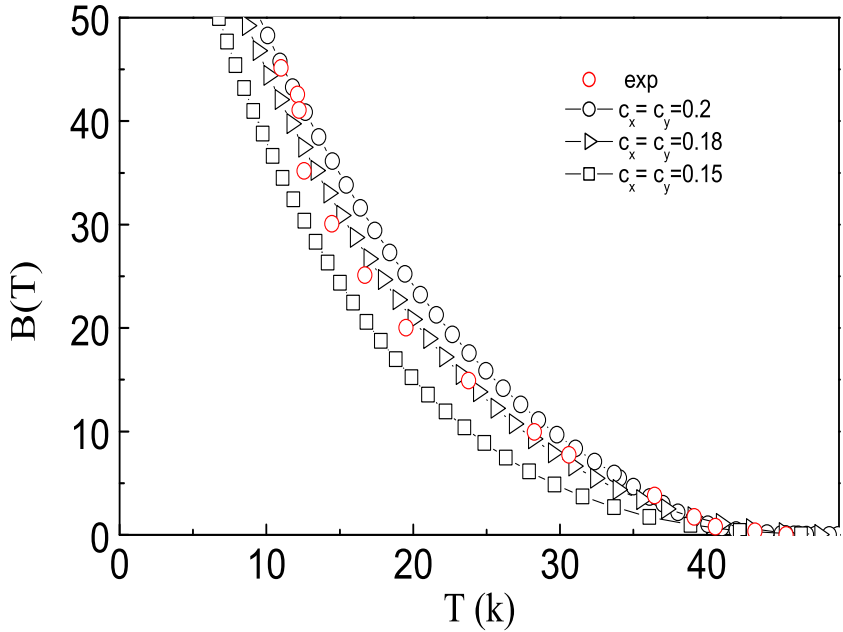}
\caption{\label{Figure Graph1}  Vortex-lattice melting lines for
$Nd(O_{1-x}F_{x})FeAs$ with  magnetic field parallel to c-axis.}
\end{figure}

We now turn to another typical iron-based layered superconductor
$Nd(O_{1-x}F_{x})FeAs$(1111-type)  \cite{1111,Ying,1111b,1111c}. We
take parameters from  \cite{1111b}, $T_{c}=49$K,
$\lambda_{\bot}=200nm$, $\xi_{\bot}=2.3nm$, $H_{c2}^{c}(T=0)=95$T,
$H_{c2}^{ab}(T=0)=220$T, $\gamma^{2}=25$ . The separation  between
the FeAs planes is given by  twice of the lattice parameter $s=2c
\approx  17  \AA $ \cite{Kamihara,Ying,dft}.

Similarly, we first calculate the melting lines for the fields
parallel to c-axis, as shown in Fig. 4. Since the melting line has
not been explicitly reported, the data at which the resistance
reaches the small percentage ($0.5\%$) of the normal-state
resistance in Ref. \cite{1111b} is regarded the approximated melting
one here, which is also collected in Fig. 4. Interestingly, the
theoretical melting line with $c=0.18$ is consistent with the
experimental data in a considerable wide temperature range. The
discrepancy between experiment and theory becomes larger at high
magnetic fields and low temperatures. A trend of temperature
independent melting data is shown  at high magnetic fields in
experiments,  implying that the  dimensional crossover occurs at
this regime. If the vortex system is of two dimensional nature at
high magnetic fields, the melting line is almost independent of the
temperature\cite{blatterG},

Next, We  calculate the melting lines when the fields perpendicular
to c-axis, which are shown in Fig. 5. For perpendicular fields,
three different groups of Linderman numbers
($C_{x}=0.1,C_{y}=0.1,C_{y}=0.075$;
$C_{x}=0.2,C_{y}=0.2,C_{y}=0.15$; $C_{x}=0.3,C_{y}=0.3,C_{y}=0.225$)
are used to locate the melting lines. A good collapse of the melting
lines in two directions can  be also achieved if setting the ratio $
c_x/c_y \approx 1.33$, nearly the same as that in 122 type.

We also  investigate  the intrinsic layer pinning effect on the
vortex melting in $Nd(O_{1-x}F_{x})FeAs$ sample. The results are
exhibited in Fig. 6. Surprisingly, we observe  an intersection
between  two melting lines with two transverse directions  for
typical Lindemann numbers, which is not shown in the above 122 type.
Obviously, a collapse of these two curves could  not be achieved by
setting any ration $ c_x/c_y $. It is implied that the intermediate
smectic phase \cite{Calson,huchen,nie} may exist in
$Nd(O_{1-x}F_{x})FeAs$. However, its identification is beyond the
phenomenological Lindemann theory. It is interesting to note that,
when the exponentially weak intrinsic pinning was taken into account
for a weakly anisotropic $Ba(Fe_{1-x}Co_{x})_{2}As_{2}$, the vortex
melting behavior is only changed slightly, but for the strongly
anisotropic $Nd(O_{1-x}F_{x})FeAs$, the intrinsic pinning  would
play a much more significant role.

\begin{figure}
\includegraphics[width=1.0 \textwidth,height=8.cm]{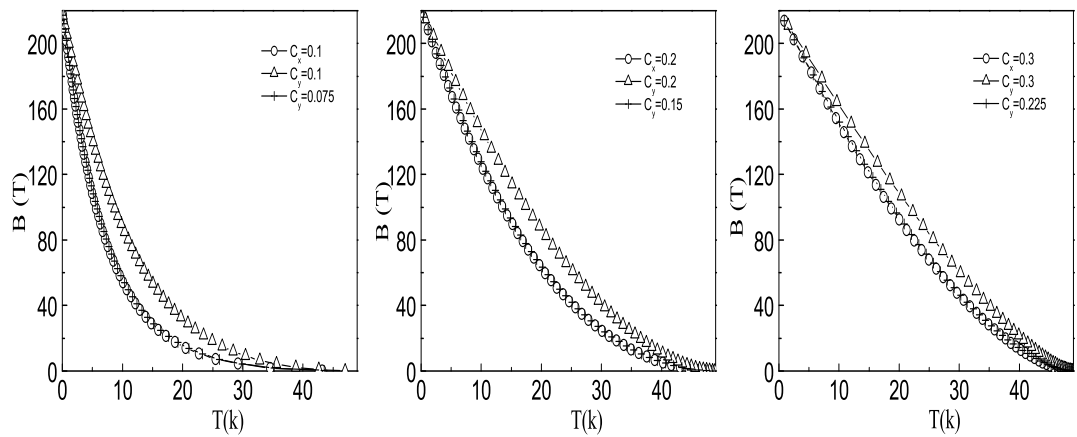}
\caption{\label{Figure Graph1}  Vortex-lattice melting lines for
$Nd(O_{1-x}F_{x})FeAs$ with  magnetic field perpendicular to c-axis
(without intrinsic layer pinning). From left to right,
$C_{x}=0.1,C_{y}=0.1,C_{y}=0.075$; $C_{x}=0.2,C_{y}=0.2,C_{y}=0.15$;
$C_{x}=0.3,C_{y}=0.3,C_{y}=0.225$.}
\end{figure}

\begin{figure}
\includegraphics[width=0.8 \textwidth,height=10.cm]{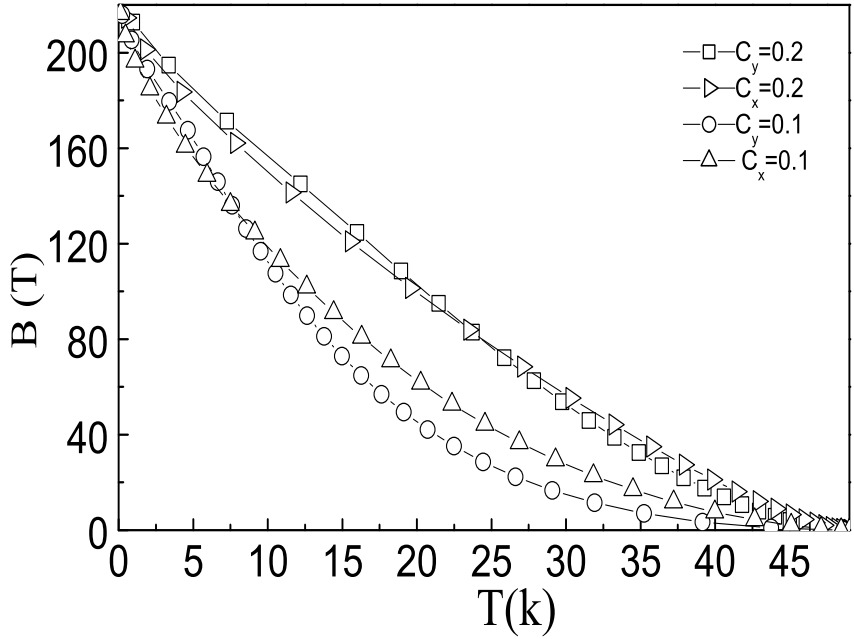}
\caption{\label{Figure Graph1}  Vortex-lattice melting lines for
$Nd(O_{1-x}F_{x})FeAs$ with   magnetic field perpendicular to c-axis
(with  intrinsic layer pinning).}
\end{figure}

\section{Summary}

In terms of the elastic theory, we have studied the thermal
fluctuations in two typical iron-based superconductors
 $Ba(Fe_{1-x}Co_{x})_{2}As_{2}$ and  $Nd(O_{1-x}F_{x})FeAs$
for magnetic fields parallel and perpendicular to the anisotropy
axis. Using the parameters of these superconductors from
experiments, the melting lines with various Lindemann numbers are
composed. Interestingly,  we can derive the melting line observed in
both $Ba(Fe_{1-x}Co_{x})_{2}As_{2}$ \cite{122} and
$Nd(O_{1-x}F_{x})FeAs$ \cite{1111b} for magnetic field parallel to
the $c-$axis. Neglecting the layer pinning effect, it is found that
thermal fluctuations  normalized by vortex separations in the two
transverse directions for parallel field  are proportional to each
other, similar to those observed in cuprate
superconductors\cite{huchen,nie}. Interestingly, the intrinsic layer
pinning play much more important role on the vortex melting in
$Nd(O_{1-x}F_{x})FeAs$ than in $Ba(Fe_{1-x}Co_{x})_{2}As_{2}$.  More
experimental works are motivated to confirm our prediction by
comparing the melting dada with the interpolation one presented
here. The present prediction may in turn provide a guide or
reference to locate the melting line in future experiments.

\section{Acknowledgements}

This work was supported by National Natural Science Foundation of
China under Grant Nos. 10774128(QHC)  and 10804098(QMN), PCSIRT
(Grant No. IRT0754) in University in China,  National Basic Research
Program of China (Grant Nos. 2006CB601003 and 2009CB929104),
Zhejiang Provincial Natural Science Foundation under Grant No.
Z7080203, and Program for Innovative Research  Team  in Zhejiang
Normal University,

\end{document}